\documentclass[10pt, conference, letterpaper]{IEEEtran}

\IEEEoverridecommandlockouts

\usepackage{wrapfig}
\usepackage{graphicx}
  \usepackage{caption}
  \usepackage [autostyle, english = american]{csquotes}
  \usepackage{subcaption}
  \usepackage[table,xcdraw]{xcolor}
  \usepackage{hyperref}
  \usepackage{comment}
  \usepackage{lipsum}
 \usepackage{dblfloatfix}
  
\usepackage{cite}
\usepackage{amsmath,amssymb,amsfonts}
\usepackage{algorithmic}
\usepackage{algorithm,balance}
\usepackage[applemac]{inputenc} 
\newcommand{\RN}[1]{%
  \textup{\uppercase\expandafter{\romannumeral#1}}%
}
\newcommand*\sepline{%
   \begin{center}
     \rule[1ex]{0.9\textwidth}{.4pt}
   \end{center}}
\usepackage{cite}
\usepackage{amsmath,amssymb,amsfonts}
\usepackage{textcomp}
\usepackage{hyperref}
\usepackage{academicons}
\usepackage{tikz,xcolor,hyperref}

\definecolor{lime}{HTML}{A6CE39}
\DeclareRobustCommand{\orcidicon}{
	\begin{tikzpicture}
	\draw[lime, fill=lime] (0,0) 
	circle [radius=0.16] 
	node[white] {{\fontfamily{qag}\selectfont \tiny ID}};
	\draw[white, fill=white] (-0.0625,0.095) 
	circle [radius=0.007];
	\end{tikzpicture}
	\hspace{-2mm}
}

\foreach \x in {A, ..., Z}{\expandafter\xdef\csname orcid\x\endcsname{\noexpand\href{https://orcid.org/\csname orcidauthor\x\endcsname}
			{\noexpand\orcidicon}}
}

\usepackage{xcolor}
\def\BibTeX{{\rm B\kern-.05em{\sc i\kern-.025em b}\kern-.08em
    T\kern-.1667em\lower.7ex\hbox{E}\kern-.125emX}}

\begin{document}

\title{ Gradient Ascent Algorithm for Enhancing Secrecy Rate in Wireless Communications for Smart Grid}

\author{Neji~Mensi\orcidA{} ,~\IEEEmembership{Graduate Student Member,~IEEE} ,  Danda~B.~Rawat\orcidB{} ,~\IEEEmembership{Senior Member,~IEEE,}\\ 
and Elyes~Balti\orcidA{} ~\IEEEmembership{Graduate Student Member,~IEEE}
\thanks{This work was supported in part by the US NSF under grants CNS 1650831.}
\thanks{
Neji Mensi and Danda B. Rawat are with the Department
of Electrical Engineering and Computer Science, Howard University, Washington, DC, 20059 USA. Contact e-mail: 
danda.rawat@howard.edu.}
\thanks{Elyes Balti is with the Wireless Networking and
Communications Group, Department of Electrical and Computer Engineering,
The University of Texas at Austin, Austin, TX 78712 USA 
}
}

\maketitle

\begin{abstract}
The emerging Internet of Things (IoT) and bidirectional communications in smart grid are expected to improve smart grid capabilities and electricity management.  Because of massive number of IoT devices in smart grid, size of the data to be transmitted increases, that demands a high data rate to meet the real-time smart grid communications requirements. Sub-6 GHz, millimeter-wave (mmWave) technologies, and massive multiple-input multiple-output (MIMO) technologies can meet high data rate demands. However, IoT enabled smart grid is still subject to various security challenges such as eavesdropping, where attackers attempt to overhear the transmitted signals and the jamming attack, where the attacker perturbs the received signals at the receiver. In this paper, our goal is to investigate jamming and eavesdropping attacks while improving secrecy capacity for smart grid communications. Specifically, we propose to employ a hybrid beamforming design for wireless communications in smart energy grid. {\color{black}In previous works, the secrecy capacity is increased by randomly augmenting the source power or setting the system combiners. Unlike state-of-the-art, we design and evaluate the Gradient Ascent algorithm to search for the best combiners/waveform that maximizes the secrecy capacity in smart grid communications.} We also study two different optimization scenarios by considering both fixed and variable transmit power. Numerical results are used for performance evaluation and supporting our formal analysis. 
\end{abstract}


\section{Introduction}
\subsection{Background and Literature Review}
The traditional grid systems were designed to deliver electricity to customers while adopting simple one-way directional interaction.  This approach makes it challenging for the grid to respond to the plethora and rising energy demands. Therefore, the smart grid was introduced as a solution to offer bidirectional information and electricity flow between utilities and their customers. It is a developing network of communicated and connected devices that helps to control power consumption and moderate the electricity needs of a wide range of customers \cite{mavroeidakos2020threat, gunduz2020cyber}. Smart home area network (HAN) is part of a smart grid network, where the different smart devices cooperate through an energy management system \cite{han}. HAN can help adjusting the operation-schedules of the connected energy devices, mitigating the electricity demand on the grid during the pick hours, and decreasing customers' energy bills.

The data to be transmitted in IoT enabled smart grid is huge because of massive number of connected smart grid devices and operations of those devices in the system.  IoT enabled smart grid provides a possibility for a large number of objects/things to interact, communicate, and collaborate using wireless communications. It enables smart grids to manage network efficiently and guarantee different operations, measurements,  and maintenance activities \cite{survIptSG}. Due to the exponentially increasing number of IoT devices in smart grid, the traditional microwave band is not enough and incapable to support smart grid communication requirements. Fortunately, Sub-6 GHz and millimeter-wave (mmWave) bands have been regarded as promising solutions to address the aforementioned constraints and to offer a highly reliable low latency communications \cite{mmW}. Because of the critical nature of the energy grid, the smart  grid is vulnerable to various security attacks, which may lead to power outage \cite{survIptSG, liu2017combating, rawat2015detection}. Furthermore, information in smart grid communication is tied to private information through HAN.

\textcolor{black}{In wireless communication systems, the eavesdropping attack is one of the most severe cyber threats where eavesdropper attempts to passively overhear/listen to the data exchanged between authorized devices in smart grid. It is a critical attack, especially if the intercepted signal is classified or if the disclosing of the secret message may influence the smart grid operations \cite{gunduz2020cyber, rawat2018smart}. To confront this challenge,  security techniques are mainly based on application and transport layers encryption. 
}
The encryption approaches suffer from two major problems; the first issue is related to the attacker's computational capabilities. If an eavesdropper has powerful computational hardware/algorithmic resources, he may be able to compromise the encryption protocols. The second dilemma is the fact that the distribution and the management of cryptography keys are challenging, especially in networks that are characterized by arbitrary access and highly distributed like in smart grid (e.g., \cite{kong2020review}). Therefore, physical layer security (PLS)  has been emerged as an alternative solution to encryption and as an emerging security engineering that mitigates passive attacks. The main idea of PLS is to take advantage of the physical layer properties such as the radio channels characteristics. One of the PLS techniques is employing a friendly jammer, whose main function is to jam and perturb the eavesdropper's link \cite{myJam}. Thereby, the signal-to-interference-plus-noise ratio (SINR) at the eavesdropper will be decreased/deteriorated which resulting in hindering the eavesdropper from overhearing the communications. This technique improves the average secrecy capacity and significantly protects the communications between the legitimate nodes. Another promising techniques is the  physical layer key generation, which was introduced as a PLS scheme \cite{PLSKey1, PLSkey2}.It is still facing some technical challenges such as co-located attacks, high bit disagreement ratio in low signal-to-noise ratio regimes, and high temporal correlation \cite{PLSkey3Mag}.

\subsection{Background}
The beamforming strategy is based on employing an antenna array, where the phase shifters can be adjusted to offer high beamforming gain and tocompensate for the path loss and multipath fading. Since mmWave signals are characterized by a small wavelength,  it becomes easier to increase the number of antennas per array within a small area which necessarily help improving the antenna gain \cite{arraySize} and serve multiple users.  Though, a large number of antennas comes with technical challenges where it necessitates a larger number of phase shifters. Due to the power consumption, fabrication cost,  and hardware limitations, it is not practical to assign a dedicated radio frequency (RF) chain controller to each phase shifter. With the aim at reducing the number of RF chains, the hybrid-transceivers design is introduced. It is composed of an analog part that is equipped with analog phase shifter networks in the RF domain and a digital part in the baseband. For a given transmitter, the digital precoder  processes data streams, then it up-converts its output with the digital-to-analog converters (DACs) to the RF  chain, which will be mapped to the transmitting antennas through the analog precoder. At the the receiver's front-end, the received signal is processed through the analog combiners and RF chains, respectively. Then, it down-converts its output to the baseband using the analog-to-digital converters (ADCs) to create the data stream and decode the received message\cite{hybridMag,hybrid1,hybrid2}. 

\subsection{Our Contribution}
The connected devices in the smart grid may be subject to an eavesdropping and jamming attack at the same time which could directly affect the quality of service (QoS) in smart grid or cause power outage.  In this paper, we leverage a machine learning technique (named Gradient Ascent) to enhance security while considering a hybrid beamforming design to cope and mitigate the effect of both jamming and eavesdropping attacks. We focus on searching for the optimum analog/digital precoders at the access point or base station (BS) and the combiners at the receivers to increase the secrecy capacity with the aid of the  Gradient Ascent approach.
To the best of the authors' knowledge, unlike the previous work in the literature,  this is the first attempt to maximize the secrecy capacity using the Gradient Ascent approach for wireless communications in smart grid network while considering eavesdropping and jamming attack, multi-user/multi-device\footnote{Without loss of generality, end user and end devices are interchangeably used throughout this paper for smart grid communications.}  multiple-input multiple-output (MU-MIMO),  and the scenarios of fixed/variable transmission power. 
Specifically, our contributions include:
\begin{enumerate}
\item[$\bullet$] Protect MU-MIMO from jamming and eavesdropping attacks while considering two different scenarios: Sub-6 GHz and mmWave systems for smart grid communications.
  \item[$\bullet$] Design a low-complexity hybrid precoding   with the aid of Gradient Ascent approach to maximize the secrecy capacity for fixed source power in smart grid communications.
   \item[$\bullet$] Design a second low-complexity hybrid precoding to reach a desirable secrecy capacity while adapting the source power to different values. 
\end{enumerate}
\subsection{Paper Structure}
The rest of the paper is organized as follows:  Section \RN{2} presents the attack scenario and the system model. Section \RN{3} describes and studies the optimization procedure of the secrecy capacity.  Then, Section \RN{4} assesses the performance of the security strategy and the energy efficiency, based on numerical results. Finally, we outline our conclusion in Section \RN{5}.
\section{System Model}
\subsection{Notation}
For the purpose of simplicity and organization, we utilize the following notations: $||\cdot||_{2}$ is the 2-norm, $|\cdot|$ is the absolute value, $||\cdot||_{F}$  is the Frobenius norm, $\mathbb{E}[\cdot]$ is the expectation, $(\cdot)^{T}$ is the matrix transpose, and  $(\cdot)^{*}$ is the Hermitian operator.
Lower case regular letter represents a running variable, upper case regular letter represents a scaler number, lower case bold letter represents a vector, upper case bold letter represents a matrix.
\subsection{Communication Network and Attack Model}
\begin{figure}
  \centering
\includegraphics[ width=\linewidth]{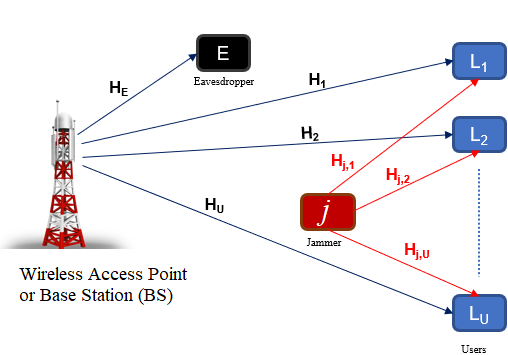}
\caption{System model with legitimate smart grid devices/users and attackers (jammer and eavesdropper)}
\label{sm1}
\end{figure}

 We are considering a general scenario of wireless communications in smart grid system, as shown in Fig. \ref{sm1}, where BS or access point communications with the  $L_u$ ($u\in\{1,2,...,U\}$) smart grid devices (and vice versa). The BS is equipped with $N_t$ antennas and  each smart grid IoT device is equipped with $N_r$ antennas. In this scenario, an eavesdropper $E$ (with $N_E$ receiving antennas) is overhearing the transmitted signal from the BS to the legitimate receivers. In addition to the eavesdropping, the legitimate smart grid IoT receivers are suffering from a jamming attack caused by malicious jammer $j$. The jammer $j$ is equipped with $N_{j}$ transmitting antennas. Its purpose is perturbing and minimizing the SINR at the legitimate receiver device. We assumed that the jammer is not attacking the eavesdropper.

\subsection{Communication Channel Model}
In this paper, {\color{black} we consider a Sub-6 GHz/mmWave system. The channels form the BS to the legitimate users $\bf{H}_u$, the channels from 
} 
the jammer to legitimate users $\bf{H}_{j,u}$ ($u\in\{1,2,...,U\}$), and from the BS to the eavesdropper $\bf{H}_{E}$, are assumed to follow a Geometric channel model based on clusters and number of rays per cluster \cite{algMMW}
\begin{equation}
H= \sqrt{\frac{N_R N_T}{N_{cl}N_{ray}}}\sum_{i=1}^{N_{cl}}\sum_{j=1}^{N_{ray}}\beta_{i,j}\textbf{a}_{R}(\phi_{i,j}^{a},\theta_{i,j}^{a})\textbf{a}_{T}^{*}(\phi_{i,j}^{d},\theta_{i,j}^{d}),
\end{equation}
where $N_{cl}$ and $N_{ray}$ are, respectively, the number of clusters and the number of rays. $N_R$  denotes the number of antennas at the receiver, where $N_R\in\{N_r,N_{_E}\}$. $N_T$ refers to the number of transmitting antennas, where $N_T\in\{N_t,N_j\}$.  $\beta_{i,j}$ is the complex gain of the $i^{th}$ cluster in the $j^{th}$
ray. The array response and steering vectors at the receiver $R$ (where $R\in\{u^{th}$ user, eavesdropper $E\}$)  and the transmitter $T$ (where $T\in\{$BS, jammer $j\}$) are $\textbf{a}_{R}(\phi_{i,j},\theta_{i,j})$ and $\textbf{a}_{T}^{*}(\phi_{i,j},\theta_{i,j})$, respectively. The angles of departure (AoD) and the angles of arrival (AoA) are denoted by $\phi_{i,j}^{d},\theta_{i,j}^{d}$ and $\phi_{i,j}^{a},\theta_{i,j}^{a}$, respectively. Where  $\phi_{i,j}^{d}$ and  $\phi_{i,j}^{a}$ in the azimuth plane and $\theta_{i,j}^{d}$ and  $\theta_{i,j}^{a}$ in the elevation plane.

The received signal (as in the \cite{li2017hybrid} for different problem)  at the $u^{th}$ receiver can be expressed as
\begin{equation}
\begin{aligned}
{ y_{u}} =& \textbf{w}^{*}_{u}\textbf{H}_{u}\textbf{F}_{_{RF}}\textbf{f}_{_{BB}}^{u}s_{u} + \textbf{w}^{*}_{u}\sum_{n\ne u}^{U}\textbf{H}_{u}\textbf{F}_{_{RF}}\textbf{f}_{_{BB}}^{n}s_{n} \\&+ \textbf{w}^{*}_{u}\textbf{H}_{j,u}\textbf{f}_{_{RF}}^{j}s_{j,u} + \textbf{w}^{*}_{u}\textbf{n}_{u},
\end{aligned}
\end{equation}
The received signal at the eavesdropper \textit{E} which is overhearing the legitimate communications is expressed as  
\begin{equation}
\begin{aligned}
{ y_{_{E}}}& = \textbf{w}^{*}_{_E}\sum_{u=1}^{U}\textbf{H}_{_E}\textbf{F}_{_{RF}}\textbf{f}_{_{BB}}^{u}s_{u}+ \textbf{w}^{*}_{_E}\textbf{n}_{_E},
\end{aligned}
\end{equation}
where $\textbf{w}_{u}$ is  the combiner vector at the $u^{th}$ user with dimension $N_{r}\times 1$. $\textbf{w}_{_E}$ is  the combiner vector at the eavesdropper $E$ with dimension $N_{_E}\times 1$. $\textbf{H}_{u}, \textbf{H}_{_E}$, and $\textbf{H}_{j,u}$ are respectively, the channel matrix from BS to $u^{th}$ legitimate user with dimensions $N_{r} \times N_{t}$, from BS to $E$ with dimensions $N_{_E} \times N_{t}$, and from jammer to $u^{th}$ user with dimensions $N_{r} \times N_{j}$. $\textbf{F}_{_{RF}}$ is the analog precoders at the BS with $N_t\times U$, where $\textbf{F}_{_{RF}}=[\textbf{f}_{_{RF}}^{1},\textbf{f}_{_{RF}}^{2},...,\textbf{f}_{_{RF}}^{U}]$ ( $\textbf{f}_{_{RF}}^{u}$ is the analog precoder intended to the $u^{th}$ user with dimension $N_{t} \times 1$, $u\in\{1,2,...,U\}$). $\textbf{F}_{_{BB}}$ is the digital precoder matrix at the BS with dimension $U \times U$, where  $\textbf{F}_{_{BB}}=[\textbf{f}_{_{BB}}^{1},\textbf{f}_{_{BB}}^{2},...,\textbf{f}_{_{BB}}^{U}]$ ($\textbf{f}_{_{BB}}^{u}$ is $U \times 1$ vector that defines the digital precoder intended to the $u^{th}$ user). The precoder at the jammer is denoted by $\textbf{f}_{_{RF}}^{j}$ with dimension $N_{j} \times 1$.  $s_{u}$ is the information symbol sent by the BS to $u^{th}$ user.  $s_{j,u}$ is the information symbol sent by the jammer to $u^{th}$ user. Finally, $\textbf{n}_{u}$ (with dimension $N_r\times 1$) and $\textbf{n}_{_E}$ ($N_{_E}\times 1$) are ,respectively, the zero mean additive white Gaussian noise (AWGN) at the $u^{th}$ user and the eavesdropper with variance $\sigma_{u}^2$ and $\sigma_{_E}^2$.
{\color{black} 
We note that the second term in Eq. (2) refers to the interference at the $u^{th}$ user/device due to the signals sent from the BS to the rest of legitimate users ($U-1$ users).} 

The SINR at the  $u^{th}$ legitimate receiver device/user is expressed as 
\begin{equation}
\begin{aligned}
{ \gamma_{u}}& = \frac{\frac{P_{b}}{U} |\textbf{w}^{*}_{u}\textbf{H}_{u}\textbf{F}_{_{RF}}\textbf{f}_{_{BB}}^{u}|^2}{\sigma_{u}^2 + \frac{P_{j}}{U} |\textbf{w}^{*}_{u}\textbf{H}_{j,u}\textbf{f}_{_{RF}}^{j}|^2 +\frac{P_{b}}{U}\sum_{n\ne u}^{U}|\textbf{w}^{*}_{u}\textbf{H}_{u}\textbf{F}_{_{RF}}\textbf{f}_{_{BB}}^{n}|^2},
\end{aligned}
\end{equation}

where $P_{b}$ and $P_{j}$ are the BS and the jammer transmission power, respectively. $\mathbb{E}[\textbf{ss}^{*}]=\frac{P_b}{U}\textbf{I}_{U}$, such that \textbf{s} = $[s_{_1},s_{_2},...,s_{_U}]^{T}$ is $U \times 1$ vector that defines the transmitted symbols sent by the BS and $\textbf{I}_{U}$ is the identity matrix with dimension $U \times U$. 
  $\mathbb{E}[\textbf{s}_j (\textbf{s}_j)^{*}]=\frac{ P_{j}}{N_j}\textbf{I}_{N_j}$, such that $\textbf{s}_j$ = $[s_{_{j,1}},s_{_{j,2}},...,s_{_{j,U}}]^{T}$ and $\textbf{I}_{N_j}$ is the identity matrix with dimension $N_j \times N_j$.\\
Then, we can define the achievable rate at the $u^{th}$ user by the Eq. (\ref{Cu}) on the top of the next page.

\begin{figure*}
\begin{equation}
\begin{aligned}
{ C_{u}}& = \log_2 \left( 1+ \frac{\frac{P_{b}}{U} |\textbf{w}^{*}_{u}\textbf{H}_{u}\textbf{F}_{_{RF}}\textbf{f}_{_{BB}}^{u}|^2}{\sigma_{u}^2 + \frac{P_{j}}{U} |\textbf{w}^{*}_{u}\textbf{H}_{j,u}\textbf{f}_{_{RF}}^{j}|^2 +\frac{P_{b}}{U}\sum_{n\ne u}^{U}|\textbf{w}^{*}_{u}\textbf{H}_{u}\textbf{F}_{_{RF}}\textbf{f}_{_{BB}}^{n}|^2}\right). \label{Cu}
\end{aligned}
\end{equation}
\sepline
\end{figure*}

We assume that the jammer intends to deteriorate the legitimate users only. Therefore, while overhearing the signal sent to the $u^{th}$ user, the received SINR at the eavesdropper is not impacted by the jammer and the eavesdropper's SINR is expressed as 
\begin{equation}
\begin{aligned}
{ \gamma_{_{E}}}& = \frac{\frac{P_{b}}{U} |\textbf{w}^{*}_{_E}\textbf{H}_{_E}\textbf{F}_{_{RF}}\textbf{f}_{_{BB}}^{u}|^2}{\sigma_{_E}^{2} +\frac{P_{b}}{U}\sum_{n\ne u}^{U}|\textbf{w}^{*}_{_E}\textbf{H}_{_E}\textbf{F}_{_{RF}}\textbf{f}_{_{BB}}^{n}|^2}.
\end{aligned}
\end{equation}

Then, we can express the achievable rate for the eavesdropper as 
\begin{equation}
\begin{aligned}
{ C_{_E}}& = \log_2 \left ( 1 +\frac{\frac{P_{b}}{U} |\textbf{w}^{*}_{_E}\textbf{H}_{_E}\textbf{F}_{_{RF}}\textbf{f}_{_{BB}}^{u}|^2}{\sigma_{_E}^{2} +\frac{P_{b}}{U} \sum_{n\ne u}^{U}|\textbf{w}^{*}_{_E}\textbf{H}_{_E}\textbf{F}_{_{RF}}\textbf{f}_{_{BB}}^{n}|^2} \right )
\label{Ce}
\end{aligned}
\end{equation}

Finally, the secrecy rate $C_{s}$ can be defined as follows
\begin{equation}
    C_{s}=\max[(C_{u}-C_{_E}),0]. \label{Cs1}
\end{equation} 
{\color{black}where $C_{_E}$ will be the max\{$C_{_1}$,...,$C_{_M}$\} if there are $M$ eavesdroppers.} 

\section{Optimization procedure of Secrecy Capacity}
\begin{figure}[ht]
     \centering
     \begin{subfigure}[t]{0.4\textwidth}
        \centering
         \includegraphics[width=7.1cm,  height=3.1cm]{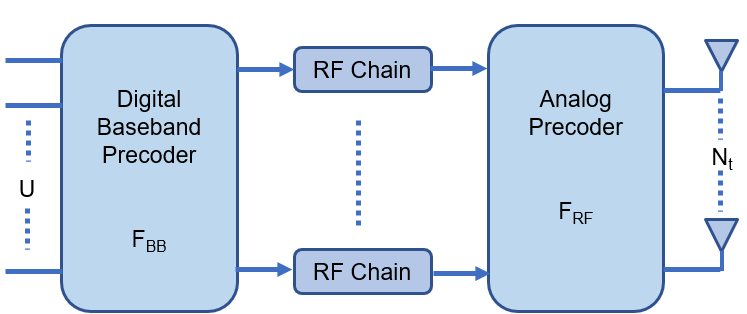}
              \vspace{+.3cm}
         \caption{MmWave MIMO hybrid design }
     \end{subfigure}
     \hfill
     \vspace{+.5cm}
     \begin{subfigure}[t]{0.4\textwidth}
         \includegraphics[width=.63\textwidth]{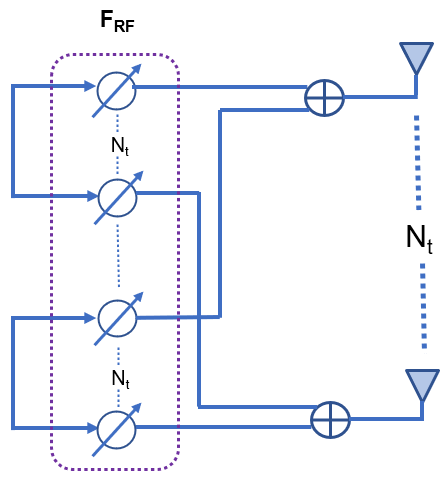}
                       \vspace{+.3cm}
         \caption{Fully Connected Analog Precoder Design}
         \vspace{+.3cm}
     \end{subfigure}
     \caption{Hybrid Beamforming architecture for wireless communication in smart energy grid. }
     \label{fig2}
\end{figure}

Our objective is to maximize the secrecy capacity $C_{s}$ for a hybrid beamforming architecture (Fig. \ref{fig2}(a)). We assume that the analog precoder is fully connected, as shown in Fig. \ref{fig2}(b). In other words, we are going to search for the best combination of three key parameters: the combiners at the users ($\textbf{w}_1,\textbf{w}_2,...,\textbf{w}_{U}$), the analog precoders $\textbf{F}_{_{RF}}$ at the BS, and the digital baseband precoder $\textbf{F}_{_{BB}}$ at the BS. 

To perform the optimization procedure, we are going to divide the maximization problem into two different stages. In the first one, we will solve for single-user MIMO (SU MIMO) analog beamforming scenario. In this stage, we aim at maximizing the received power at the $u^{th}$ user, minimizing the interference coming from the jammer, and mitigating the eavesdropping impact.   Then, in the second stage, we will consider the MU-MIMO and utilize the optimized analog precoder matrix $\textbf{F}_{_{RF}}$ and the combiner $\textbf{w}_{u}$ of each user to compute the digital precoder matrix  $\textbf{F}_{_{BB}}$ using 3 different precoding techniques: Maximum Ratio Transmission (MRT), Minimum Mean Square Error (MMSE), and Zero-Forcing (ZF). 

 \subsection{SU-MIMO Analog Beamforming}
 In this stage, we aim at maximizing the secrecy capacity of SU while ignoring the interference coming from the other legitimate users. In other words, we will maximize the received power $|\textbf{w}^{*}_{u}\textbf{H}_{u}\textbf{f}_{_{RF}}^{u}|^2$ at the $u^{th}$ user, minimize the interference power coming from the jammer $|\textbf{w}^{*}_{u}\textbf{H}_{j,u}\textbf{f}_{_{RF}}^{j}|^2$, and mitigate the eavesdropping impact on the communication confidentiality. \\
 We can reformulate the received signal at the $u^{th}$ user for the SU scenario as follows
\begin{equation}
\begin{aligned}
 y_{u}^{_{SU}}& = \textbf{w}^{*}_{u}\textbf{H}_{u}\textbf{f}_{_{RF}}^{u}s_{u} + \textbf{w}^{*}_{u}\textbf{H}_{j,u}\textbf{f}_{_{RF}}^{j}s_{j,u} + \textbf{w}^{*}_{u}\textbf{n}_{u}.
\end{aligned}
\end{equation}
The received signal at the eavesdropper \textit{E} while listening to the signal sent to the $u^{th}$ user is given by
\begin{equation}
\begin{aligned}
 y_{_{E}}^{_{SU}}& = \textbf{w}^{*}_{_E}\textbf{H}_{_E}\textbf{f}_{_{RF}}^{u}s_{u}+ \textbf{w}^{*}_{_E}\textbf{n}_{_E}.
\end{aligned}
\end{equation}

Hence, the received SINR at the $u^{th}$ user is defined by
\begin{equation}
\begin{aligned}
 \gamma_{u}^{_{SU}}& = \frac{\frac{P_{b}}{U} |\textbf{w}^{*}_{u}\textbf{H}_{u}\textbf{f}_{_{RF}}^{u}|^2}{\sigma_{u}^2 + \frac{P_{j}}{U} |\textbf{w}^{*}_{u}\textbf{H}_{j,u}\textbf{f}_{_{RF}}^{j}|^2}.
\end{aligned}
\end{equation}
The received SNR at the eavesdropper is given by
\begin{equation}
\begin{aligned}
 \gamma_{_{E}}^{_{SU}}& = \frac{\frac{P_{b}}{U} |\textbf{w}^{*}_{_E}\textbf{H}_{_E}\textbf{f}_{_{RF}}^{u}|^2}{\sigma_{_E}^{2}}.
\end{aligned}
\end{equation}
Thereby, the ergodic capacity at the $u^{th}$ user and the eavesdropper is defined by
\begin{equation}
\begin{aligned}
 C_{u}^{_{SU}}& = \log_2 \left ( 1 +   \frac{\frac{P_{b}}{U} |\textbf{w}^{*}_{u}\textbf{H}_{u}\textbf{f}_{_{RF}}^{u}|^2}{\sigma_{u}^2 + \frac{P_{j}}{U} |\textbf{w}^{*}_{u}\textbf{H}_{j,u}\textbf{f}_{_{RF}}^{j}|^2} \right ).
\end{aligned}
\end{equation}

\begin{equation}
\begin{aligned}
 C_{_{E}}^{_{SU}}& = \log_2 \left ( 1 +  \frac{\frac{P_{b}}{U} |\textbf{w}^{*}_{_E}\textbf{H}_{_E}\textbf{f}_{_{RF}}^{u}|^2}{\sigma_{_E}^{2}} \right ).
\end{aligned}
\end{equation}

 Therefore, we modify the secrecy capacity giving in Eq.(\ref{Cs1}) while ignoring the interference coming from other users. Hence, the new secrecy capacity is denoted by $C_{s}^{_{SU}}$ and defined as follows
 \begin{equation}
 \begin{aligned}
 C_{s}^{_{SU}}=&\max[(C_{u}^{_{SU}}-C_{_{E}}^{_{SU}}),0].
 \label{Cs2}
 \end{aligned}
\end{equation} 
This first stage, searches for the best values of $\textbf{w}_{u}$ and $\textbf{f}_{_{RF}}^{u}$ that maximize the secrecy capacity (our cost function) defined in (\ref{Cs2}). The optimization process needs to satisfy the following constraints: the unit-norm and the Constant Amplitude (CA) constraints.

We can formulate the optimization problem as follows
\begin{equation}\label{maxrate}
\begin{split}
\mathcal{P}: \max\limits_{\textbf{w}_{u},\textbf{f}_{_{RF}}^{u}, \textbf{f}_{_{RF}}^{j}}& C_{s}^{_{SU}}
\end{split}
\end{equation}
\begin{equation}\label{c1}
\text{subject to}~~|\textbf{f}_{_{RF}}^{u}| =1,  |\textbf{w}_{u}| =  1
\end{equation}
\begin{equation}\label{c2}
~~~~~~~~~~~~~~~~~~~~~~~~~~~~\textbf{w}_{u} \in\mathcal{G}^{N_r} ~\text{and}~ \textbf{f}_{_{RF}}^{u},  \in\mathcal{G}^{N_t}
\end{equation}
\begin{equation}\label{c3}
~~~~~~~~~~~~~~\textbf{w}_{u} = \frac{\textbf{w}_{u}}{\sqrt{N_r}|\mathbf{w}_{u}|},
\end{equation}
\begin{equation}\label{c4}
~~~~~~~~~~~~~~\textbf{f}_{_{RF}}^{u} = \frac{\textbf{f}_{_{RF}}^{u}}{\sqrt{N_t}|\textbf{f}_{_{RF}}^{u}|},
\end{equation}
where Eq. (\ref{c1}) is the unit-norm constraint and Eqs. (\ref{c3}, \ref{c4}) are the CA constraints. \begin{math}\mathcal{G}^{N_t}\end{math} and 
\begin{math}\mathcal{G}^{N_r}\end{math} are, respectively, the subspace of the CA constraints of dimensions $N_t$ and $N_r$.
We need to compute the gradients with respect to $\textbf{w}_{u}$ and $\textbf{f}_{_{RF}}^{u}$, then project the solutions onto the subspace of the CA constraints. We chose only those two parameters to compute the gradients since we have no control on the combiner $\textbf{w}_{_E}$ at the eavesdropper or the precoder $\textbf{f}_{_{RF}}^{j}$ at the jammer.
Now, we will start by deriving the cost function for $\textbf{w}_{u}$, which is expressed as shown in Eq.(\ref{dW}) on the top of the next page, where $\Psi_{j,u}=|(\textbf{w}^{*}_{u}\textbf{H}_{j,u}\textbf{f}_{_{RF}}^{j})|^2$ and $\Psi_{u}=|(\textbf{w}^{*}_{u}\textbf{H}_{u}\textbf{f}_{_{RF}}^{u})|^2$.

\begin{figure*}
\begin{equation}
\begin{aligned}
\nabla_{\textbf{w}_{u}^{*}}C_{s}^{_{SU}} &=\frac{\partial C_{s}^{_{SU}}}{\partial\textbf{w}_{u}^{*}}= \left[ \frac{\partial C_{u}^{_{SU}}}{\partial\textbf{w}_{u}^{*}}-\frac{\partial C_{_{E}}^{_{SU}}}{\partial\textbf{w}_{u}^{*}} \right]=\frac{\partial C_{u}^{_{SU}}}{\partial\textbf{w}_{u}^{*}}\\
&=\frac{\textbf{w}_{u}\sigma_{u}^{2} + \frac{P_{j}}{U}\textbf{H}_{j,u}\textbf{f}_{_{RF}}^{j}(\textbf{f}_{_{RF}}^{j})^{*}\textbf{H}_{j,u}^{*}\textbf{w}_{u} +\frac{P_{b}}{U}\textbf{H}_{u}\textbf{f}_{_{RF}}^{u}(\textbf{f}_{_{RF}}^{u})^{*}\textbf{H}_{u}^{*}\textbf{w}_{u}}{\sigma_{u}^{2} + \frac{P_{j}}{U}\Psi_{j,u}+\frac{P_{b}}{U}\Psi_{u} }
-\frac{\textbf{w}_{u}\sigma^2 + \frac{P_{j}}{U}\textbf{H}_{j,u}\textbf{f}_{_{RF}}^{j}(\textbf{f}_{_{RF}}^{j})^{*}\textbf{H}_{j,u}^{*}\textbf{w}_{u} }{\sigma_{u}^{2} + \frac{P_{j}}{U} \Psi_{j,u}},\label{dW}
\end{aligned}
\end{equation}
\end{figure*}

As a second step, we compute the gradient of $\textbf{f}_{_{RF}}^{u}$, which is given by

\begin{equation}
\begin{aligned}
\nabla_{(\textbf{f}_{_{RF}}^{u})^{*}}C_{s}^{_{SU}}& =\frac{\partial C_{s}^{_{SU}}}{\partial(\textbf{f}_{_{RF}}^{u})^{*}}= \left[ \frac{\partial C_{u}^{_{SU}}}{\partial(\textbf{f}_{_{RF}}^{u})^{*}}-\frac{\partial C_{_{E}}^{_{SU}}}{\partial(\textbf{f}_{_{RF}}^{u})^{*}} \right].
\end{aligned}
\end{equation}
Hence, we can first compute the partial derivative of the rate at the legitimate receiver node
\begin{equation}
\begin{aligned}
\frac{\partial C_{u}^{_{SU}}}{(\textbf{f}_{_{RF}}^{u})^{*}} &=
\frac{\frac{P_{b}}{U}\textbf{H}_{u}^{*}\textbf{w}_{u}\textbf{w}_{u}^{*}\textbf{H}_{u}\textbf{f}_{_{RF}}^{u}}{\sigma_{u}^{2} + P_{j} \Psi_{j,u} + \frac{P_{b}}{U} \Psi_{u}}.
\end{aligned}
\end{equation}
Then, we  derive the partial derivative of the rate at the eavesdropper
\begin{equation}
\begin{aligned}
\frac{\partial C_{_{E}}^{_{SU}}}{(\textbf{f}_{_{RF}}^{u})^{*}} &=
\frac{\frac{P_{b}}{U}\textbf{H}_{_E}^{*}\textbf{w}_{_E}\textbf{w}_{_E}^{*}\textbf{H}_{_E}\textbf{f}_{_{RF}}^{u}}{\sigma_{_E}^{2} + \frac{P_{b}}{U}\Psi_{_E}},
\end{aligned}
\end{equation}
where $\Psi_{E}=|(\textbf{w}^{*}_{_E}\textbf{H}_{_E}\textbf{f}_{_{RF}}^{u})|^2$.\\
Thereby, we can express the gradient  of $\textbf{f}_{_{RF}}^{u}$ by
\begin{equation}
\begin{aligned}
\nabla_{(\textbf{f}_{_{RF}}^{u})^{*}}C_{s}^{_{SU}}=\frac{\frac{P_{b}}{U}\textbf{H}_{u}^{*}\textbf{w}_{u}\textbf{w}_{u}^{*}\textbf{H}_{u}\textbf{f}_{_{RF}}^{u}}{\sigma_{u}^{2} + P_{j} \Psi_{j,u} + \frac{P_{b}}{U} \Psi_{u}} -\frac{\frac{P_{b}}{U}\textbf{H}_{_E}^{*}\textbf{w}_{_E}\textbf{w}_{_E}^{*}\textbf{H}_{_E}\textbf{f}_{_{RF}}^{u}}{\sigma_{_E}^{2} +  \frac{P_{b}}{U}\Psi_{_E}}
\end{aligned}
\end{equation}

\subsection{MU-MIMO Digital Beamforming}
The previous stage of SU analog beamforming optimization is repeated $U$ times. In other word, the analog optimization scheme should be processed for each user. Thereby, as an output, we obtain 
the best combination of all combiners at all the users $\textbf{w}_1,\textbf{w}_2,...,\textbf{w}_{U}$ and the analog precoder  $\textbf{F}_{_{RF}}$. Now, we turn our attention to the digital precoding design for multi-user where we compute the digital baseband precoder $\textbf{F}_{_{BB}}$. We start by computing the effective channel $\textbf{h}_{eff}^{u}$ of each user $u$ with dimension $1\times U$ using effective feedback techniques \cite{feedBack1,feedBack2}.\\  $\textbf{h}_{eff}^{u}$ is defined as follows
\begin{equation}
   \textbf{h}_{eff}^{u}=\textbf{w}_{u}^{*}\textbf{H}_{u}\textbf{F}_{_{RF}}. 
\end{equation}

Next, each user will feedback its effective channel to the BS. The BS builds channel $\textbf{H}$, where $\textbf{H}=[(\textbf{h}_{eff}^{1})^{T},(\textbf{h}_{eff}^{2})^{T},...,(\textbf{h}_{eff}^{U})^{T}]$.

 To design the digital precoding matrix $\textbf{F}_{_{BB}}$, we use three different precoding filters \cite{heath}. The first one is ZF, where the digital precoding matrix is denoted by $\textbf{F}_{_{BB}}^{zf}$ and defined by
 \begin{equation}
   \textbf{F}_{_{BB}}^{zf}=\textbf{H}^{*}(\textbf{H}\textbf{H}^{*})^{-1}.
\end{equation}
The second filter is the MMSE, where the digital precoding matrix is denoted and given by 
 \begin{equation}
   \textbf{F}_{_{BB}}^{mmse}=\textbf{H}^{*}\left(\textbf{H}\textbf{H}^{*}+\frac{U}{SNR}\textbf{I}_U\right)^{-1},
\end{equation}
where $\textbf{I}$ is the identity matrix with dimension $U \times U$.
The third filter is the MRT, where it is denoted by $\textbf{F}_{_{BB}}^{mrt}$ and defined as follows 
 \begin{equation}
   \textbf{F}_{_{BB}}^{mrt}=\textbf{H}^{*}.
\end{equation}
We remind that  $\textbf{F}_{_{BB}}=[\textbf{f}_{_{BB}}^{1},\textbf{f}_{_{BB}}^{2},...,\textbf{f}_{_{BB}}^{U}]$, where $\textbf{f}_{_{BB}}^{u}$ is the digital precoder of the $u^{th}$ user. $\textbf{f}_{_{BB}}^{u}$ needs to be normalized as follows
 \begin{equation}
   \textbf{f}_{_{BB}}^{u}=\frac{\textbf{f}_{_{BB}}^{u}}{||\textbf{F}_{_{RF}}\textbf{f}_{_{BB}}^{u}||_F}, u=1,2,...,U
\end{equation}
\subsection{Performance Analysis}
\subsubsection{Achievable Average Secrecy Capacity}
The BS serves all $U$ users simultaneously. Therefore, in addition to the interference caused by the jammer, each user $u$ suffers from interference of the other users' streams. Hence, by referring to Eqs. (\ref{Cu}, \ref{Ce}, and \ref{Cs1}), we can write the achievable average secrecy capacity as shown on Eq. (31) on the top of the next page.
 \begin{figure*}
 \begin{equation}
 \begin{aligned}
   C_{s}=\max\left[\left\{\log_{2}\left( 1+ \frac{\frac{P}{U}|\textbf{w}_{u}^{*}\textbf{H}_{u}\textbf{F}_{_{RF}}\textbf{f}_{_{BB}}^{u}|^2}{\sigma_{u}^2+\frac{P_j}{U}|\textbf{w}_{u}^{*}\textbf{H}_{j,u}\textbf{f}_{_{RF}}^{j}|^2+\frac{P_b}{U}\sum_{n \ne u}^{U}|\textbf{w}_{u}^{*}\textbf{H}_{u}\textbf{F}_{_{RF}}\textbf{f}_{_{BB}}^{n}|^2}   \right) \right. \right.~~~~~~~~ \\- \left.\left.\log_2 \left( 1 + \frac{\frac{P_{b}}{U} |\textbf{w}^{*}_{_E}\textbf{H}_{_E}\textbf{F}_{_{RF}}\textbf{f}_{_{BB}}^{u}|^2}{\sigma_{_E}^{2}+\frac{P_{b}}{U} \sum_{n\ne u}^{U}|\textbf{w}^{*}_{_E}\textbf{H}_{_E}\textbf{F}_{_{RF}}\textbf{f}_{_{BB}}^{n}|^2} \right )\right\},0\right].
\end{aligned}
\end{equation}
\sepline
\end{figure*}
\subsubsection{Energy Efficiency}
We note by $\Xi$ the  energy efficiency per user $u$ as the ratio between the achievable secrecy rate per user and the total power consumption, where $\Xi$ is expressed in bits/Hz/Joule. It is given by \cite{algMMW}
\begin{equation}\label{EE}
    \Xi=\frac{C_{u}}{P_{common}+N_{_{RF}}P_{_{RF}}+N_{t}P_{PA}+N_{PS}},
\end{equation}

where $P_{common}$ is the common power of the transmitter ($P_b$ in our study scenario), $N_{_{RF}}$ and $P_{_{RF}}$ are the number and power of RF chain, respectively. 
We note that an RF chain consists of  down-converter, low-noise amplifier, and DAC/ADC. $P_{PA}$ and $P_{PS}$ are respectively, the power of the amplifier and the power of the phase shifter. $N_{PS}$ is given by
$ 
\begin{cases} {N_{PS}}=
N_{t}N_{_{RF}} ~~~~ \textrm{Fully-Connected (our proposed system)}\\
N_{PS}=N_{t} ~~~~~~~~~~ \textrm{Partially-Connected}
\end{cases}
$

\subsection{Optimization Algorithm}
Now, after presenting the two optimization stages: SU-MIMO analog beamforming and the MU-MIMO digital beamforming, {\color{black} we will summarize the optimization steps for fixed/variable source power $P_{b}$.  
Algorithm 1 (SU-MIMO analog beamforming with fixed source power $P_{b}$) and Algorithm 2 (MU-MIMO digital beamforming with fixed source power $P_{b}$)  search for the maximum secrecy capacity while maintaining the source power $P_{b}$ fixed. This is very efficient in the case where the power is limited and we want to achieve the maximum secrecy capacity while using limited power resources. Algorithm 3 (SU-MIMO analog beamforming with variable source power $P_{b}$) and Algorithm 4 (MU-MIMO digital beamforming with variable source variable $P_{b}$) optimize  with respect to a secrecy capacity target $\zeta$ required by the QoS. 
} 
In this second approach,  we search for the maximum possible secrecy capacity while adapting $P_{b}$. If we achieve $\zeta$, then the algorithm stops, otherwise it adapts $P_{b}$ while the power cannot exceed a certain threshold $\mu$ (to avoid infinite loop and unrealistic power  consumption). 
We note that $\delta$ is the gradient algorithm step size, $\epsilon$ is the convergence criterion, and $\kappa$ is the power adaptation rate.

\begin{algorithm}
 \caption{\\SU-MIMO analog beamforming with fixed source power ($P_{b}$)}\label{algo1}
 \begin{algorithmic}[1]
 \renewcommand{\algorithmicrequire}{\textbf{Input:}}
 \renewcommand{\algorithmicensure}{\textbf{Output:}}
 \renewcommand{\algorithmiccomment}{$\triangleright~$}
 \REQUIRE $\delta$ , $\epsilon$, $\textbf{H}_{u}$,~$\textbf{H}_{_E}$,~$\textbf{w}_{_E}$,~$\textbf{H}_{j,u}$,~$\textbf{f}_{_{RF}}^{j}$
 \ENSURE $\textbf{w}_{u}$,~$\textbf{f}_{_{RF}}^{u}$\\
  \STATE \textbf{Initialize} $\textbf{w}_{u} = \textbf{w}_{u}^{(0)},~\textbf{f}_{_{RF}}^{u} = \textbf{f}_{_{RF},(0)}^{u}$
  \WHILE{$|C_{s,(n+1)}^{_{SU}} - C_{s,(n)}^{_{SU}}| > \epsilon$} 
    \STATE $C_{s,(n)}^{_{SU}} \gets C_{s,(n+1)}^{_{SU}}$
  \STATE $\textbf{w}_{u}^{(n+1)} \gets \textbf{w}_{u}^{(n)}+\delta \nabla_{\textbf{w}^{*}_{u}}C_{s,(n)}^{_{SU}}$
  \STATE $\textbf{w}_{u}^{(n+1)} \gets \frac{\textbf{w}_{u}^{(n+1)}}{\|\textbf{w}_{u}^{(n+1)}\|_2}$
  \COMMENT{Unit-norm constraint}
  \STATE $\textbf{w}_{u}^{(n+1)} \gets \frac{\textbf{w}_{u}^{(n+1)}}{\sqrt{N_R}|\textbf{w}_{u}^{(n+1)}|}$
  \COMMENT{CA constraint}
  \STATE $\textbf{f}_{_{RF},(n+1)}^{u} \gets \textbf{f}_{_{RF},n}^{u}+\delta \nabla_{(\textbf{f}_{_{RF}}^{u})^{*}}C_{s,(n)}^{_{SU}}$
  \STATE $\textbf{f}_{_{RF},(n+1)}^{u} \gets \frac{\textbf{f}_{_{RF},(n+1)}^{u}}{\|\textbf{f}_{_{RF},(n+1)}^{u}\|_2}$
   \STATE $\textbf{f}_{_{RF},(n+1)}^{u} \gets \frac{\textbf{f}_{_{RF},(n+1)}^{u}}{\sqrt{N_t}|\textbf{f}_{_{RF},(n+1)}^{u}|}$
    \IF{$C_{s,(n+1)}^{_{SU}}<C_{s,(n)}^{_{SU}}$} \STATE {Adapt $\delta$} \ENDIF
  \ENDWHILE
 \RETURN $\textbf{w}_{u}$,~$\textbf{f}_{_{RF}}^{u}$
 \end{algorithmic} 
 \end{algorithm}
 
  \begin{algorithm} 
 \caption{\\MU-MIMO digital beamforming with fixed source power ($P_{b}$)}\label{algo2}
 \begin{algorithmic}[1]
 \renewcommand{\algorithmicrequire}{\textbf{Input:}}
 \renewcommand{\algorithmicensure}{\textbf{Output:}}
 \renewcommand{\algorithmiccomment}{$\triangleright~$}
  \FORALL{$u=1,2,...,U$} 
    \STATE Call \textbf{Algorithm 1}
    \STATE return $\textbf{f}_{_{RF}}^{u}$, $\textbf{w}_{u}$\ENDFOR
 \FORALL{$u=1,2,...,U$} 
    \STATE $\textbf{h}_{eff}^{u}=\textbf{w}_{u}^{*}\textbf{H}_{u}\textbf{F}_{_{RF}}$
    \COMMENT{Compute the effective channel of  the $u^{th}$ user}
    \STATE \textbf{return} $\textbf{H}$
         \COMMENT{Effective channel of all users}
    \ENDFOR    
\STATE $\textbf{F}_{_{BB}}^{zf}=\textbf{H}^{*}(\textbf{H}\textbf{H}^{*})^{-1}~~~~~~~~~~~~~~~~$
     \COMMENT{Option1: ZF}
        \STATE $ \textbf{F}_{_{BB}}^{mmse}=\textbf{H}^{*}\left(\textbf{H}\textbf{H}^{*}+\frac{U}{SNR}\textbf{I}\right)^{-1}~~~$
     \COMMENT{Option 2: MMSE} 
        \STATE $\textbf{F}_{_{BB}}^{mrt}=\textbf{H}^{*}~~~~~~~~~~~~~~~~~~~~~~~~~~$
     \COMMENT{Option 3: MRT}  
 \FORALL{$u=1,2,...,U$} 
    \STATE $\textbf{f}_{_{BB}}^{u}=\frac{\textbf{f}_{_{BB}}^{u}}{||\textbf{F}_{_{RF}}\textbf{f}_{_{BB}}^{u}||_F}, u=1,2,...,U$
    \COMMENT{Normalize the digital precoder of the $u^{th}$ user}
    \ENDFOR 
 \STATE \textbf{Compute} and \textbf{return} $C_s$    
 \end{algorithmic} 
 \end{algorithm}

 \begin{algorithm}[]
 \caption{\\SU-MIMO analog beamforming with variable source power ($P_{b}$)}\label{algo3}
 \begin{algorithmic}[1]
 \renewcommand{\algorithmicrequire}{\textbf{Input:}}
 \renewcommand{\algorithmicensure}{\textbf{Output:}}
 \renewcommand{\algorithmiccomment}{$\triangleright~$}
 \REQUIRE $\delta$ , $\epsilon$, $\textbf{H}_{u}$,~$\textbf{H}_{_E}$,~$\textbf{w}_{_E}$,~$\textbf{H}_{j,u}$,~$\textbf{f}_{_{RF}}^{j}$
 \ENSURE $\textbf{w}_{u}$,~$\textbf{f}_{_{RF}}^{u}$\\
  \STATE \textbf{Initialize} $\textbf{w}_{u} = \textbf{w}_{u}^{(0)},~\textbf{f}_{_{RF}}^{u} = \textbf{f}_{_{RF},(0)}^{u}$
  \WHILE{$C_{s,(n+1)}^{_{SU}} < \zeta$  and $P_{b} \leq \mu$} 
  \STATE Call \textbf{Algorithm 1} from line \textbf{2} to line \textbf{13}
  \STATE  $P_{b} \gets P_{b}(1+\kappa)~~~~~~$
   \COMMENT{Adapt$ ~P_{b}$ with rate $\kappa$}
  \ENDWHILE 
 \RETURN $\textbf{w}_{u}$,~$\textbf{f}_{_{RF}}^{u}$
 \end{algorithmic} 
 \end{algorithm}
 
   \begin{algorithm}[]
 \caption{\\MU-MIMO digital beamforming with variable source power ($P_{b}$)}\label{algo4}
 \begin{algorithmic}[1]
 \renewcommand{\algorithmicrequire}{\textbf{Input:}}
 \renewcommand{\algorithmicensure}{\textbf{Output:}}
 \renewcommand{\algorithmiccomment}{$\triangleright~$}
  \FORALL{$u=1,2,...,U$} 
    \STATE Call \textbf{Algorithm 2}
    \STATE return $\textbf{f}_{_{RF}}^{u}$, $\textbf{w}_{u}$\ENDFOR
 \FORALL{$u=1,2,...,U$} 
    \STATE $\textbf{h}_{eff}^{u}=\textbf{w}_{u}^{*}\textbf{H}_{u}\textbf{F}_{_{RF}}$
    \COMMENT{Compute the effective channel of  the $u^{th}$ user}
    \STATE \textbf{return} $\textbf{H}$
         \COMMENT{Effective channel of all users}
    \ENDFOR    
\STATE $\textbf{F}_{_{BB}}^{zf}=\textbf{H}^{*}(\textbf{H}\textbf{H}^{*})^{-1}~~~~~~~~~~~~~~~~~$
     \COMMENT{Option1: ZF}
        \STATE $ \textbf{F}_{_{BB}}^{mmse}=\textbf{H}^{*}\left(\textbf{H}\textbf{H}^{*}+\frac{U}{SNR}\textbf{I}\right)^{-1}~~~~$
     \COMMENT{Option 2: MMSE} 
\STATE $\textbf{F}_{_{BB}}^{mrt}=\textbf{H}^{*}~~~~~~~~~~~~~~~~~~~~~~~~~~~$
     \COMMENT{Option 3: MRT}  
 \FORALL{$u=1,2,...,U$} 
    \STATE $\textbf{f}_{_{BB}}^{u}=\frac{\textbf{f}_{_{BB}}^{u}}{||\textbf{F}_{_{RF}}\textbf{f}_{_{BB}}^{u}||_F}, u=1,2,...,U$
    \COMMENT{Normalize the digital precoder of the $u^{th}$ user}
    \ENDFOR 
 \STATE \textbf{Compute} and \textbf{return} $C_s$    
 \end{algorithmic} 
 \end{algorithm}
 
 \section{Numerical Results and Discussions}

\begin{table}[H]
\caption{System Parameters}
\centering
\begin{tabular}{|c|c|c|}
\hline
\rowcolor[HTML]{FFCC67} 
Parameters                             & MmWaves & Sub-6 GHz \\ \hline
\begin{tabular}[c]{@{}c@{}}Number of antennas at the receivers:\\ $N_t$ and $N_{_E}$\end{tabular} & 4       & 2         \\ \hline
\rowcolor[HTML]{FFCC67} 
Number of  clusters                    & 4       & 10        \\ \hline
Number of rays per cluster             & 15      & 20        \\ \hline
\rowcolor[HTML]{FFCC67} 
Angular spread                         & $10^{o}$       & $10^{o}$         \\ \hline
$\delta$                                 & $10^{-1}$     & $10^{-1}$        \\ \hline
\rowcolor[HTML]{FFCC67} 
$\epsilon$                                &  $10^{-7}$     & $10^{-7}$      \\ \hline
$\kappa$                                  & $10^{-2}$     & $10^{-2}$          \\ \hline
\rowcolor[HTML]{FFCC67} 
Number of Monte Carlo Simulation loops & 1000    & 1000      \\ \hline
\end{tabular}
\end{table}

In this section, we will examine the performance of the proposed security scheme.  To guarantee the convergence stability and to adjust any possible perturbation during the learning process, the step size parameter $\delta$ is initialized to 0.1 and it is divided by 2 whenever a perturbation occurs during the optimization cycle of the cost function. The channel matrices $\textbf{H}_{u}$, $\textbf {H}_{_E}$, $\textbf{H}_{j,u}$, the combiners $\textbf{w}_{u}$, $\textbf{w}_{_E}$ and the precoders $\textbf{f}_{u}$, and $\textbf{f}_{_{RF}}^{j}$ are initialized to the complex Gaussian distribution. Note that the parameters mentioned in Tableau 1 are fixed  for all simulation figures.

 \begin{figure}[]
\includegraphics[ width=\linewidth]{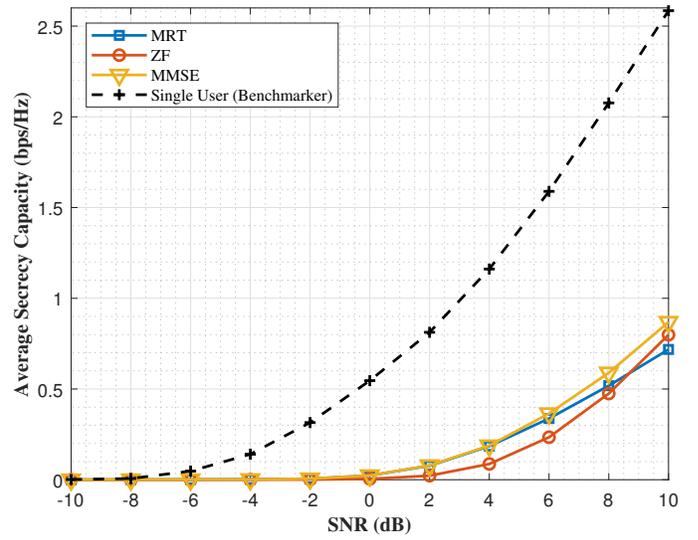}
\caption{ The average secrecy capacity for all $U$ users with respect to the SNR for the mmWave scenario. Setting parameters are: $P_{j}$= -20 dB, $N_{tj}$=16, $N_t$=64}
\end{figure}

We commence the system performance analysis by studying the average secrecy capacity. 
 Fig. 3, shows the average secrecy capacity for all $U$ users with respect to the SNR under the mmWave scenario.  We assumed that the jamming power $P_{j}$=-20 dB, the number of antennas at the jammers $N_{j}$=16, the number of antennas at the BS $N_{t}$=64, and the number of users is $U$=5. We studied three different filters MRT, ZF, and MMSE. 
 {\color{black}
 As shown, the average secrecy capacity is very low for SNR less than 2 dB for all three filters. Therefore, the source power from the BS needs to be increased since receiver is suffering from jamming, interference coming from other users, and eavesdropping.
 }
 When the SNR starts to increase, we remark that in the range from 2 dB to 9 dB, the average secrecy starts to increase with different rates. The MMSE filter provides higher average secrecy than ZF and MRT filters, while the  MRT filter outperforms the ZF filter. For the high SNR region, the ZF filter outperforms the MRT filter and provides average secrecy of 0.8 bps/Hz, while the MRT achieves only 0.6 bps/Hz. We computed the secrecy capacity for a SU which serves as a benchmarker. We note the secrecy for a SU is higher than the MU scenario, since SU and MU both are subject to eavesdropping and jamming attacks,  SU outperforms MU because it does not suffer from interference coming from other users.   

 \begin{figure}
\includegraphics[ width=\linewidth]{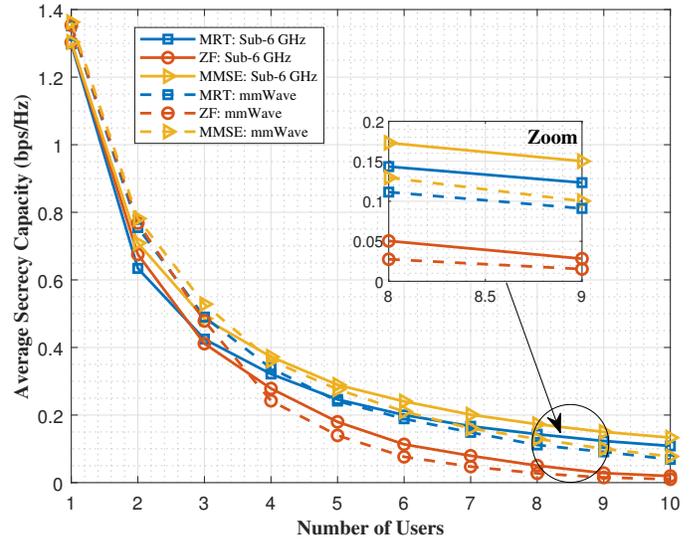}
\caption{Average secrecy capacity for Sub-6 GHz and mmWave with respect to the number of users. Setting parameters: $P_{b}$=5 dB,$N_{j}$=4, $P_j$= -10 dB, $N_t=$ 16 for Sub-6 GHz and 64 for mmWave.}
\end{figure}
Fig. 4 presents the average secrecy capacity for Sub-6 GHz and mmWave with respect to the number of users. In this plot, we set the BS power $P_{b}$ to 5 dB, the number of antennas at the jammer $N_{j}$ to 4, and the jamming power $P_j$ to -10 dB. The number of antennas at the BS is 16 for Sub-6 GHz and 64 for the mmWave scenario. We observe that the average secrecy capacity decreases as the number of users increases. It starts with 1.37 bps/Hz for a SU, then it declines to 0.175 for 10  users while using MMSE filter for the Sub-6 GHz. This is caused by the interference coming from the served users. Under the aforementioned system parameters and for 2 users or less, the robustness of both Sub-6 GHz and mmWave systems to the jamming and eavesdropping attacks are very close, even while using different precoding filters. On the other hand, when the number of users rises, Sub-6 GHz system slightly outperforms the mmWave system. Moreover, the performance of each used filter starts to be distinguished. MMSE provides higher average secrecy than both ZF and MRT.  To surmount the decline of average secrecy capacity with a large number of users,  we need to increase the number of antennas at the BS as well as $P_{b}$.

\begin{figure}
\includegraphics[ width=\linewidth]{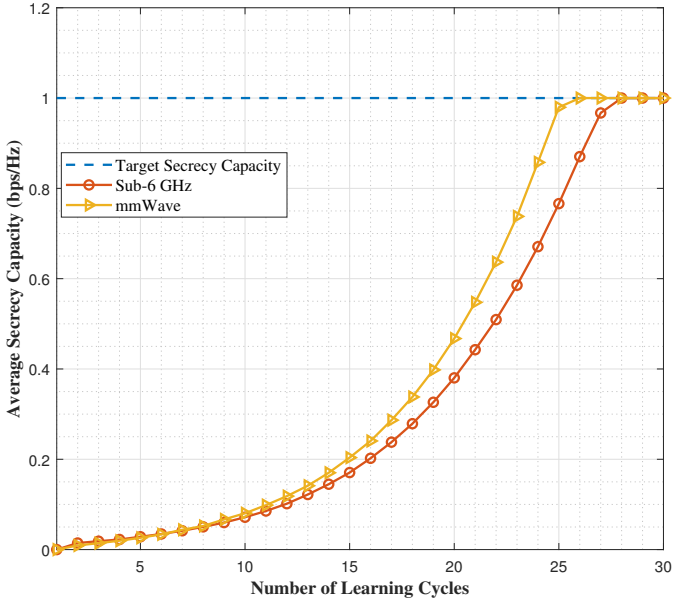}
\caption{Sub-6 GHz vs. mmWave with MMSE filter: Comparison in terms of source power $P_{b}$  and number of cycles required to reach a target secrecy capacity  $\zeta$. Setting parameters are: $U$=5, $N_{j}$=4, $P_j$=-5 dB, $N_t=$ 16 for Sub-6 GHz and 64 for mmWave.}
\end{figure}
{\color{black}
For the previous Fig. 4, we discussed the results of Algorithms 1 and 2, where $P_{b}$ is fixed. Now, we move to analyze Algorithm 3 and 4 by discussing Fig. 5.
} 
In this scenario, the source power from the BS is considered to be variable. Our goal is to achieve a target secrecy capacity noted by $\zeta$.  The algorithm searches for the maximum possible $C_{s}$ for a given $P_{b}$, if the obtained value reaches $\zeta$, the learning process will stop. If not, the source power is adapted and another learning cycle is initiated (where a cycle is referring to the learning process for a given $P_{b}$). In Fig. 5, the target secrecy capacity is $\zeta$=1 bps/Hz while $P_{b}$ starts at -10 dB.  The algorithm arrives at the required $\zeta$ after 26 cycles for mmWave system, while it requires 28 cycles for Sub-6 GHz. Regarding the source power needed to obtain $\zeta$, at the end of the entire learning process, $P_{b}$  increased from -10 dB to 11.02 dB for Sub-6 GHz while it requires only 9.6 dB in the case of mmWave. Therefore, under the condition of secrecy capacity requirement, we conclude that the learning rate (number of cycles required to reach $\zeta$) for mmWave system case is much higher than the Sub-6 GHz while it requires less source power. 

 \begin{figure}
\includegraphics[ width=\linewidth]{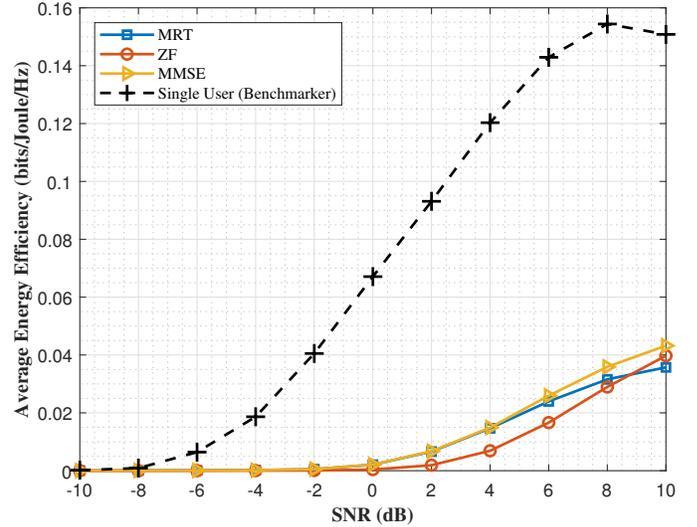} 
\caption{{\color{black}Average energy efficiency with respect to the SNR for mmWave system. Setting parameters are: $P_{j}$= -20 dB, $N_{j}$= 16, $N_t$= 64.}}
\end{figure}
After discussing the average secrecy capacity, we turn our attention to examine the energy efficiency in Fig. 6 and Fig. 7.
In both figures, we set  $P_{_{RF}}$ and $P_{PA}$ to 100 mW,  $P_{PS}$  to 10 mW \cite{algMMW}, and $U$  to 5. 
Fig. 6 emphasizes on the average energy efficiency with respect to the SNR for the mmWave system, where $P_{j}$= -20 dB, $N_{j}$= 16, and $N_t$= 64. We note that for a low SNR regime (SNR $<$ 2 dB), the energy efficiency is very low. It begins to improve with the increase of SNR. This is explained by the fact that the average secrecy capacity is low within this SNR range (refer to Eq. \ref{EE} and Fig. 4). For the high SNR regime, the average energy efficiency improves with a different rate for each used precoding filter.  At SNR=10 dB, the energy efficiency reaches 0.045 bits/Joule/Hz, 0.04 bits/Joule/Hz, and 0.035 bits/Joule/Hz for MMSE, ZF, and MRT, respectively. Moreover, the gap between the average energy efficiency for SU and MU scenarios is explained by the impact of the interference caused by the users. Therefore, as the number of served users increases, the energy efficiency decreases. 
 \begin{figure}
\includegraphics[width=\linewidth]{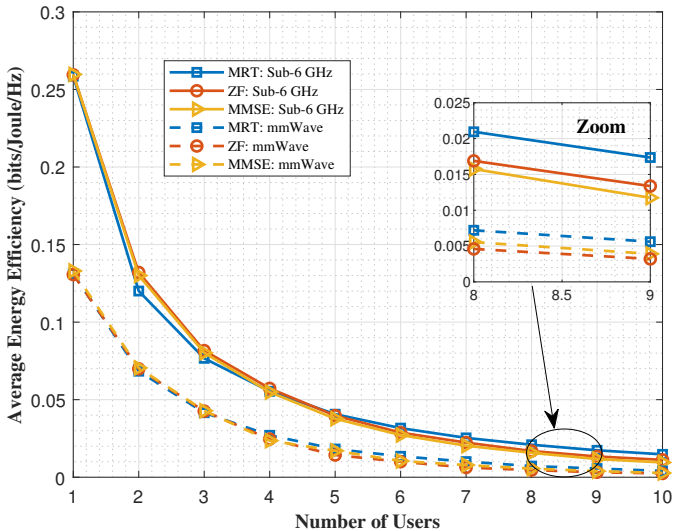} 
\caption{{\color{black}Average energy efficiency for Sub-6 GHz and mmWave with respect to the number of users. Setting parameters are: $P_{b}$=5 dB,$N_{j}$=4, $P_j$= -10 dB, $N_t=$ 16 for Sub-6 GHz and 64 for mmWave.}}
\end{figure}
In Fig. 7, we further investigate the impact of the number of users on the energy efficiency for the Sub-6 GHz and mmWave systems. We set the BS power $P_{b}$ to 5 dB, the number of antennas at the jammer $N_{j}$ to 4, and the jamming power $P_j$ to -10 dB. For $U<$ 4 users, Sub-6 GHz provides a higher average energy efficiency than the mmWave system. As the number of users increases, the performances of the two systems start to converge to similar values.

\section{Conclusion}
In this work, we discussed the PLS for an MU-MIMO communication scenario in a smart grid framework. We assumed Sub-6 GHz and mmWave systems where the legitimate receivers are subject to eavesdropping and jamming attacks.  Our goal was to optimize the hybrid precoding system with aim at improving the secrecy capacity and mitigating the impact of: jamming attack, eavesdropping attack, and interference caused by the other served users.  To perform the security scheme, we divided the optimization process into two stages. In the first stage, we considered only the SU model where we ignored the multiuser interference. Within this stage, we used a Gradient ascent approach to compute the best analog precoders and combiners that maximize the secrecy capacity. In the second stage, we minimize the multiuser interference  by computing the digital precoder matrix at the BS with 3 different filters: MMSE, ZF, and MRT. Moreover,  we proposed two different scenarios.  In the first one,  we assumed that the source power is fixed and we searched for the best hybrid precoding parameters that maximize the secrecy capacity. In the second scenario, we supposed that the QoS requires a specific secrecy capacity threshold that we need to achieve while using the optimum source power. 
We concluded that the secrecy capacity and the energy efficiency are highly sensitive to the number of users, where the Sub-6 GHz is slightly more robust than the mmWave system.
On the other hand, when we aimed at achieving a specific secrecy capacity, the mmWave system reached the secrecy target faster than Sub-6 GHz and with lower source power.

\section*{Acknowledgments}
	This work was supported in part by the US NSF under grants CNS 1650831  and by the DoD Center of Excellence in AI and Machine Learning (CoE-AIML) at Howard University under Contract Number W911NF-20-2-0277 with the U.S. Army Research Laboratory. 
 However, any opinion, finding, and conclusions
or recommendations expressed in this document are those of the authors
and should not be interpreted as necessarily representing the official
policies, either expressed or implied, of the funding agencies.

\bibliographystyle{IEEEtran}
\bibliography{bibliography}

\begin{thebibliography}{10}
\providecommand{\url}[1]{#1}
\csname url@samestyle\endcsname
\providecommand{\newblock}{\relax}
\providecommand{\bibinfo}[2]{#2}
\providecommand{\BIBentrySTDinterwordspacing}{\spaceskip=0pt\relax}
\providecommand{\BIBentryALTinterwordstretchfactor}{4}
\providecommand{\BIBentryALTinterwordspacing}{\spaceskip=\fontdimen2\font plus
\BIBentryALTinterwordstretchfactor\fontdimen3\font minus
  \fontdimen4\font\relax}
\providecommand{\BIBforeignlanguage}[2]{{%
\expandafter\ifx\csname l@#1\endcsname\relax
\typeout{** WARNING: IEEEtran.bst: No hyphenation pattern has been}%
\typeout{** loaded for the language `#1'. Using the pattern for}%
\typeout{** the default language instead.}%
\else
\language=\csname l@#1\endcsname
\fi
#2}}
\providecommand{\BIBdecl}{\relax}
\BIBdecl

\bibitem{mavroeidakos2020threat}
T.~Mavroeidakos and V.~Chaldeakis, ``{Threat Landscape of Next Generation
  IoT-Enabled Smart Grids},'' in \emph{IFIP International Conference on
  Artificial Intelligence Applications and Innovations}.\hskip 1em plus 0.5em
  minus 0.4em\relax Springer, 2020, pp. 116--127.

\bibitem{gunduz2020cyber}
M.~Z. Gunduz and R.~Das, ``Cyber-security on smart grid: Threats and potential
  solutions,'' \emph{Computer networks}, vol. 169, p. 107094, 2020.

\bibitem{han}
H.~{Albataineh}, M.~{Nijim}, and D.~{Bollampall}, ``The design of a novel smart
  home control system using smart grid based on edge and cloud computing,'' in
  \emph{2020 IEEE 8th International Conference on Smart Energy Grid Engineering
  (SEGE)}, 2020, pp. 88--91.

\bibitem{survIptSG}
F.~{Dalipi} and S.~Y. {Yayilgan}, ``Security and privacy considerations for iot
  application on smart grids: Survey and research challenges,'' in \emph{2016
  IEEE 4th International Conference on Future Internet of Things and Cloud
  Workshops (FiCloudW)}, 2016, pp. 63--68.

\bibitem{mmW}
Y.~{Yang}, Z.~{Gao}, Y.~{Ma}, B.~{Cao}, and D.~{He}, ``Machine learning
  enabling analog beam selection for concurrent transmissions in
  millimeter-wave v2v communications,'' \emph{IEEE Transactions on Vehicular
  Technology}, vol.~69, no.~8, pp. 9185--9189, 2020.

\bibitem{liu2017combating}
Y.~Liu, Y.~Zhou, and S.~Hu, ``Combating coordinated pricing cyberattack and
  energy theft in smart home cyber-physical systems,'' \emph{IEEE Transactions
  on Computer-Aided Design of Integrated Circuits and Systems}, vol.~37, no.~3,
  pp. 573--586, 2017.

\bibitem{rawat2015detection}
D.~B. Rawat and C.~Bajracharya, ``Detection of false data injection attacks in
  smart grid communication systems,'' \emph{IEEE Signal Processing Letters},
  vol.~22, no.~10, pp. 1652--1656, 2015.

\bibitem{rawat2018smart}
D.~B. Rawat and K.~Z. Ghafoor, \emph{Smart cities cybersecurity and
  privacy}.\hskip 1em plus 0.5em minus 0.4em\relax Elsevier, 2018.

\bibitem{kong2020review}
P.-Y. Kong, ``A review of quantum key distribution protocols in the perspective
  of smart grid communication security,'' \emph{IEEE Systems Journal}, 2020.

\bibitem{myJam}
N.~Mensi, D.~B. Rawat, and E.~Balti, ``{PLS for {V2I} Communications Using
  Friendly Jammer and Double kappa-mu Shadowed Fading},'' in \emph{2021 IEEE
  International Conference on Communications (IEEE ICC'21)}, Montreal, Canada,
  Jun. 2021.

\bibitem{PLSKey1}
K.~{Zeng}, ``Physical layer key generation in wireless networks: challenges and
  opportunities,'' \emph{IEEE Communications Magazine}, vol.~53, no.~6, pp.
  33--39, 2015.

\bibitem{PLSkey2}
K.~{Zeng}, D.~{Wu}, A.~{Chan}, and P.~{Mohapatra}, ``Exploiting
  multiple-antenna diversity for shared secret key generation in wireless
  networks,'' in \emph{2010 Proceedings IEEE INFOCOM}, 2010, pp. 1--9.

\bibitem{PLSkey3Mag}
L.~{Jiao}, N.~{Wang}, P.~{Wang}, A.~{Alipour-Fanid}, J.~{Tang}, and K.~{Zeng},
  ``Physical layer key generation in 5g wireless networks,'' \emph{IEEE
  Wireless Communications}, vol.~26, no.~5, pp. 48--54, 2019.

\bibitem{arraySize}
S.~{Zihir}, O.~D. {Gurbuz}, A.~{Karroy}, S.~{Raman}, and G.~M. {Rebeiz}, ``A 60
  ghz single-chip 256-element wafer-scale phased array with eirp of 45 dbm
  using sub-reticle stitching,'' in \emph{2015 IEEE Radio Frequency Integrated
  Circuits Symposium (RFIC)}, 2015, pp. 23--26.

\bibitem{hybridMag}
A.~F. {Molisch}, V.~V. {Ratnam}, S.~{Han}, Z.~{Li}, S.~L.~H. {Nguyen}, L.~{Li},
  and K.~{Haneda}, ``Hybrid beamforming for massive mimo: A survey,''
  \emph{IEEE Communications Magazine}, vol.~55, no.~9, pp. 134--141, 2017.

\bibitem{hybrid1}
A.~{Adhikary}, J.~{Nam}, J.~{Ahn}, and G.~{Caire}, ``Joint spatial division and
  multiplexing-the large-scale array regime,'' \emph{IEEE Transactions on
  Information Theory}, vol.~59, no.~10, pp. 6441--6463, 2013.

\bibitem{hybrid2}
Z.~{Chen}, X.~{Zhang}, S.~{Wang}, Y.~{Xu}, J.~{Xiong}, and X.~{Wang}, ``Bush:
  Empowering large-scale mu-mimo in wlans with hybrid beamforming,'' in
  \emph{IEEE INFOCOM 2017 - IEEE Conference on Computer Communications}, 2017,
  pp. 1--9.

\bibitem{algMMW}
X.~{Yu}, J.~{Shen}, J.~{Zhang}, and K.~B. {Letaief}, ``Alternating minimization
  algorithms for hybrid precoding in millimeter wave mimo systems,'' \emph{IEEE
  Journal of Selected Topics in Signal Processing}, vol.~10, no.~3, pp.
  485--500, 2016.

\bibitem{li2017hybrid}
A.~Li and C.~Masouros, ``{Hybrid precoding and combining design for
  millimeter-wave multi-user MIMO based on SVD},'' in \emph{2017 IEEE
  International Conference on Communications (ICC)}, 2017, pp. 1--6.

\bibitem{feedBack1}
O.~E. {Ayach}, S.~{Rajagopal}, S.~{Abu-Surra}, Z.~{Pi}, and R.~W. {Heath},
  ``Spatially sparse precoding in millimeter wave mimo systems,'' \emph{IEEE
  Transactions on Wireless Communications}, vol.~13, no.~3, pp. 1499--1513,
  2014.

\bibitem{feedBack2}
D.~J. {Love} and R.~W. {Heath}, ``Limited feedback unitary precoding for
  spatial multiplexing systems,'' \emph{IEEE Transactions on Information
  Theory}, vol.~51, no.~8, pp. 2967--2976, 2005.

\bibitem{heath}
R.~W. Heath~Jr. and A.~Lozano, \emph{Foundations of MIMO Communication}.\hskip
  1em plus 0.5em minus 0.4em\relax Cambridge University Press, 2018.

\end{thebibliography}

\begin{wrapfigure}{l}{0.2\textwidth}
\centering
\includegraphics[width=0.17\textwidth]{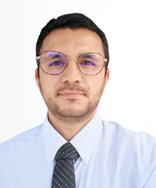}
\caption*{}
\end{wrapfigure}
\textbf{Neji Mensi} (Graduate Student Member, IEEE) received the Engineering degree in telecommunications from Ecole Nationale d'Electronique et des Telecommunications de Sfax, Tunisia, in 2016, and the M.S. degree in electrical engineering from Universite Paul Sabatier, France, in 2018. He is currently pursuing the Ph.D. degree in electrical engineering with the Department of Electrical Engineering and Computer Science, Howard University, Washington, DC, USA, under the supervision of Dr. Danda B. Rawat. During Spring 2016, he served as a Research Scholar with the University of Idaho, Moscow, ID, USA.He was a Software Engineer with Ooredoo, Tunisia, in 2017, and INTM, France, in 2019. His research focuses on physical layer security, wireless communications, and machine learning.\\

\begin{wrapfigure}{l}{0.2\textwidth}
\centering
\includegraphics[width=0.17\textwidth]{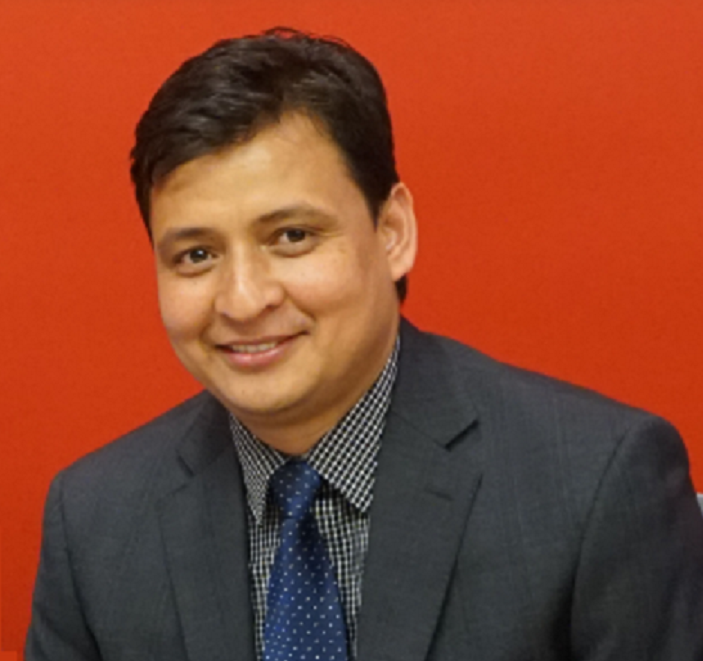}
\caption*{}
\end{wrapfigure}
\textbf{Danda B. Rawat} (Senior Member, IEEE) received the Ph.D. degree from Old Dominion University, Norfolk, VA, USA. He is a Full Professor
with the Department of Electrical Engineering \& Computer Science, the Director of the Data Science and Cybersecurity Center, DoD Center of Excellence in AI/ML, Cyber-Security and Wireless Networking Innovations Research Lab, and Graduate Cybersecurity Certificate Program, and a Graduate Program Director of Graduate CS Programs with Howard University, Washington, DC, USA. He has secured over \$16 million in research funding from the U.S. National Science Foundation (NSF), U.S. Department of Homeland Security (DHS), U.S. National Security Agency, U.S. Department of Energy, National Nuclear Security Administration, DoD and DoD Research Labs, Industry (Microsoft and Intel) and private Foundations. He has published over 200 scientific/technical articles and ten books. He is engaged in research and teaching in the areas of cybersecurity, machine learning, big data analytics and wireless networking for emerging networked systems including cyber-physical systems, Internet-of-Things, multi domain battle, smart cities, software defined systems, and vehicular networks. He is the recipient of the NSF CAREER Award in 2016, the DHS Scientific Leadership Award in 2017, the Researcher Exemplar Award 2019 and Graduate Faculty Exemplar Award 2019 from Howard University, the U.S. Air Force Research Laboratory Summer Faculty Visiting Fellowship in 2017, the Outstanding Research Faculty Award (Award for Excellence in Scholarly Activity) at GSU in 2015, the Best Paper Awards (IEEE CCNC, IEEE ICII, and BWCA), and the Outstanding Ph.D. Researcher Award in 2009. He has delivered over 20 Keynotes and invited speeches at international conferences and workshops. He has been in Organizing Committees for several IEEE flagship conferences, such as IEEE INFOCOM, IEEE CNS, IEEE ICC, and IEEE GLOBECOM. He has been serving as an Editor/Guest Editor for over 50 international journals, including the Associate Editor of IEEE TRANSACTIONS OF SERVICE COMPUTING and IEEE TRANSACTIONS OF NETWORK SCIENCE AND ENGINEERING, an Editor of IEEE INTERNET OF THINGS JOURNAL, and a Technical Editor of IEEE NETWORK. He served as a Technical Program Committee Member for several international conferences, including IEEE INFOCOM, IEEE GLOBECOM, IEEE CCNC, IEEE GreenCom, IEEE ICC, IEEE WCNC, and IEEE VTC conferences. He served as a Vice Chair of the Executive Committee of the IEEE Savannah Section from 2013 to 2017. He is an ACM Distinguished Speaker. He is a Senior Member of ACM, a member of ASEE and AAAS, and a Fellow of the Institution of Engineering and Technology.\\

\begin{wrapfigure}{l}{0.2\textwidth}
\centering
\includegraphics[width=0.17\textwidth]{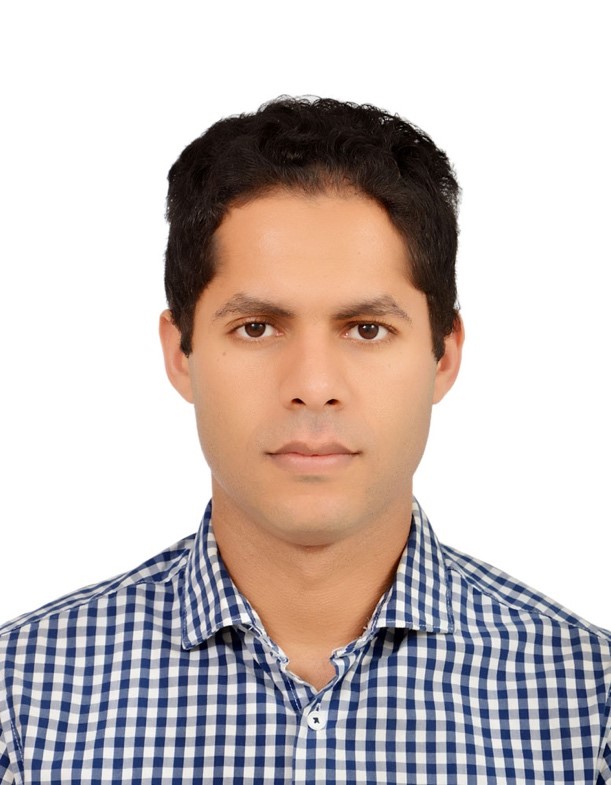}
\caption*{}
\end{wrapfigure}
\textbf{Elyes Balti} (Graduate Student Member, IEEE)received the BS and MS degrees in electrical engineering from The Ecole Superieure des Communications de Tunis (Sup'Com), Tunisia, in 2013, and The University of Idaho, ID, USA, in 2018, respectively. He is currently working toward completion of the Ph.D. program in electrical engineering at The University of Texas at Austin, TX, USA, where he is a member of the Wireless Networking and Communications Group (WNCG). He has held summer internships at Motorola Mobility, Chicago, IL, USA in 2019, Huawei, Bridgewater, NJ, USA in 2020 and Qualcomm, San Diego, CA, USA in 2021. His research interests are in wireless communications, signal processing and massive MIMO.

\end{document}